\documentclass[11pt]{article}
\pdfoutput=1

\usepackage{url}
\usepackage{graphics}
\usepackage{times}

\title{Avatar-independent scripting for real-time gesture animation}
\author{Richard Kennaway\\School of Computing Sciences,\\University of East Anglia, Norwich NR4 7TJ, U.K.}
\date{2 November 2006}

\newcommand{\citeNP}[1]{{\cite{#1}}}
\newcommand{\emdash}{---}
\newcommand{\xe}[1]{\texttt{$<$#1$>$}}
\newcommand{\signname}[1]{\textit{#1}}

\begin{document}
\maketitle

\begin{abstract}
When animation of a humanoid figure is to be generated at run-time, instead of by replaying pre-composed motion clips, some method is required of specifying the avatar's movements in a form from which the required motion data can be automatically generated.
This form must be of a more abstract nature than raw motion data:
ideally, it should be independent of the particular avatar's proportions, and both writable by hand and suitable for automatic generation from higher-level descriptions of the required actions.

We describe here the development and implementation of such a scripting language for the particular area of sign languages of the deaf, called SiGML (Signing Gesture Markup Language), based on the existing HamNoSys notation for sign languages.

We conclude by suggesting how this work may be extended to more general animation
for interactive virtual reality applications.
\end{abstract}

\section{Introduction}

In interactive applications of virtual reality, it is necessary to create animations of avatars and other objects at run-time, according to the demands of the moment.
Usually, this is done by replaying motion clips selected from a library of pre-composed motions, created by either manual keyframing or motion capture. This method suffers from several disadvantages.
\begin{enumerate}
\item
It may require a very large motion library to anticipate every possible action that may be demanded.
\item
The same movement will be played in exactly the same way every time it is used, reducing the realism of the scene.
\item
The motion data are specific to one avatar.
\end{enumerate}

It is possible to warp motion clips, for example, to adapt a motion for one avatar to another avatar, to have a walk cycle for a character follow a specified path through the world, or to smoothly blend together the end of one clip with the start of another. However, this only gives a limited amount of extra flexibility.
Inspection of some current video games indicates that at present little run-time warping or blending is done.

Run-time generation of the motion data from some higher-level description would allow for animation to be generated in a much more flexible way, in order to exactly fit the interactive situation at each moment.
In this paper we will describe an avatar-independent scripting language for posture and movement in the particular domain of sign languages of the deaf, and how we use it to automatically generate animation data in real time for any avatar.
A secondary advantage of the method is that the script for a movement is many times smaller than the motion data generated from it, which greatly reduces storage requirements.

The software~\cite{kennaway:synthetic} was developed in the ViSiCAST~\cite{Ell00,VerTF:sign:01}
and eSign~[\citeNP{eSignCVHI};
\url{http://www.sign-lang.uni-hamburg.de/eSIGN}] projects,
as one component of a system which provides
signing animations for websites or other interactive multimedia applications. (The software was not deployed until the eSign project; ViSiCAST still relied on motion-captured animations.)
A web page using this sytem will include an avatar on the page (implemented as an ActiveX control), and beside each block of text for which signing is available, a button to request that the signing be played. The web page also contains data to drive the avatar when one of these buttons is clicked, not in the form of motion data, but as text in the avatar-independent scripting language.  This is translated in real time into frames of motion data (bone rotations and facial morphs).
The data frames are passed to the avatar and played at a fixed frame rate.
A fuller account of the whole system can be found in~\cite{eSignCVHI}.
A publicly available web site employing the technology is at
\url{http://www.gebarennet.nl}\footnote{Note that the text of the web site is in Dutch, and the signing is in Dutch Sign Language. The animations will play on Internet Explorer running on Windows (2000 or XP), and require installing some software (available from the web site). There is also a video demo on the site for those without the necessary configuration. The live avatar plays rather more smoothly than the video.}.

The scripting language and its translation to motion data are the subject of the present paper.
Our concern is with the concrete geometry of the animation, in contrast to the large number of proposals for scripting languages for embodied conversational agents, which focus on much higher-level concepts. Some of these are briefly surveyed in Section~\ref{section:otherwork}.

\section{Related work}\label{section:otherwork}

There have been numerous proposals for scripting languages for embodied conversational agents, including
VHML~[\citeNP{MarS02}; \url{http://www.vhml.org}],
HumanML~[\url{http://www.humanmarkup.org}],
EMOTE~\cite{chi:emote},
GESTYLE~\cite{NooR03},
CML~\cite{Ara-etal02},
AML~\cite{Ara-etal02,PelB03},
APML~\cite{CarCBP02},
MURML~\cite{KraKW02},
MPML~\cite{PreDI04},
and STEP~\cite{Huang:STEP:02,Huang:implementation}.
A survey of some of these languages and several others can be found in \cite{Pel03},
and \cite{Pre:lifelike:04} is a book-length account of various implementations of ``social agents''.
These languages mainly address a higher level of abstraction than we are concerned with here, typical elements or attributes being
\textit{angry}, % most
\textit{alert}, % MPML
\textit{think}, % MPML
\textit{emphasize}, % GESTYLE
and \textit{semiote}. % VHML
The main purpose of most of these languages is to specify the behaviour of such agents, and especially the emotionally expressive facial and bodily movements that they should make to accompany realistic speech and interact in an engaging way with the user.
The specification of the concrete movements which perform these high-level concepts is typically seen as a mere implementation detail.
Among the proposals that include explicit description of concrete movement, emphasis has been primarily on facial animation, with body animation either ignored (for talking head applications), undeveloped, as in the most recent (2001) VHML specification, or implemented directly in terms of avatar-dependent joint rotations.

The VHD system~\cite{SanBTT99} for directing real-time virtual actors is based on motion capture and predefined postures, and the VHD++ system~\cite{PonPMTT03} is a high-level virtual reality engine for integrating lower-level technologies.

Dance notations such as Labanotation~\cite{hutchinson:labanotation} have been animated by software such as Don Henderson's
LINTER~[\url{http://www-staff.mcs.uts.edu.au/~don/pubs/led.html}], although that system does not attempt naturalistic animation of detailed avatars.
Labanotation and other dance notations present particular challenges to computer animation, as transcriptions written to be read by human dancers rely for their interpretation not only on the dancer's knowledge of dancing, but upon their creative input to flesh out intentionally incomplete transcriptions. Such creative extrapolations are beyond the intended scope of the work we are engaged in.

At the opposite extreme to these are the low-level FAPS (Facial Animation Parameters) and BAPS (Body Animation Parameters) of the MPEG-4 standard~\cite{PandzicF02}, and the H-Anim standard~[\url{http://www.h-anim.org}].
FAPS are a set of basic facial movements expressed in terms of a standard set of feature points of the face. Amounts of movement of these points are expressed in terms of a set of standard measurements of the face (e.g.~distance between the eyes). They are thus avatar-independent: the same stream of FAPS data played through different face models should produce animations that are seen as doing the same thing (as demonstrated in~\cite{EisG98}). BAPS, on the other hand, consist of joint rotations in the form of Euler angles, and are not avatar-independent. A stream of BAPS data which, played through one avatar, causes it to clap its hands, will fail to do so if played through another avatar which is identical except that its shoulders are a couple of inches further apart: the hands will fail to meet by the same distance. H-Anim is an adjunct to the VRML (Virtual Reality Modelling Language) standard defining a standard skeleton and form of motion data for humanoid animation. Like BAPS it expresses motion data in terms of avatar-dependent joint rotations.

We thus perceive a role for a method of specifying physical posture and movement in a manner which is avatar-independent, human-readable and -writable, and which allows unambiguous calculation of the exact avatar-dependent motion data for any given avatar.
Such a method can provide the lower level of implementation which must underlie all the higher-level proposals described above.
This is the purpose of the work reported here.

In the area of sign language,
Vcom3D~[\url{http://www.vcom3d.com}] publish a commercial product, Signing\-Avatar, for performing sign language on web pages. It uses an internal notation for specifying the movements of the avatar, of which details are not available.
SignSynth~\cite{grieve:signsynth} is a signing animation system based on the Stokoe notation, producing output in VRML, and also addressing the issue of translating English text to sign language.
The GESSYCA system~\cite{lebourque.sylvie:complete,lebourque.sylvie:high,gibet:high-level}
is concerned with French sign language, and of all the animation systems listed here is closest to the approach of the present paper.
It uses a notation system developed for a project called Qualgest (Qualitative Communication Gestures Specification), and uses biomechanical simulation to generate realistic human movements.

\section{HamNoSys and SiGML}

The language we have developed is called SiGML (Signing Gesture Markup Language), and was developed from HamNoSys, the Hamburg Notation System.
HamNoSys~\cite{prillwitz.leven.ea:hamnosys,hanke:hamnosys} is a notation for recording sign language gestures, developed by researchers on sign language at the IDGS (Institut f\"ur Deutsche Ge\-b\"ard\-en\-sprache) at the University of Hamburg.
It differs from other signing notations such as Stokoe~\cite{stokoe:dictionary} in that it is not specific to any one sign language, but is intended to cover all signing gestures in all sign languages.

\begin{figure}
  \centering
  \resizebox{8cm}{!}{\includegraphics{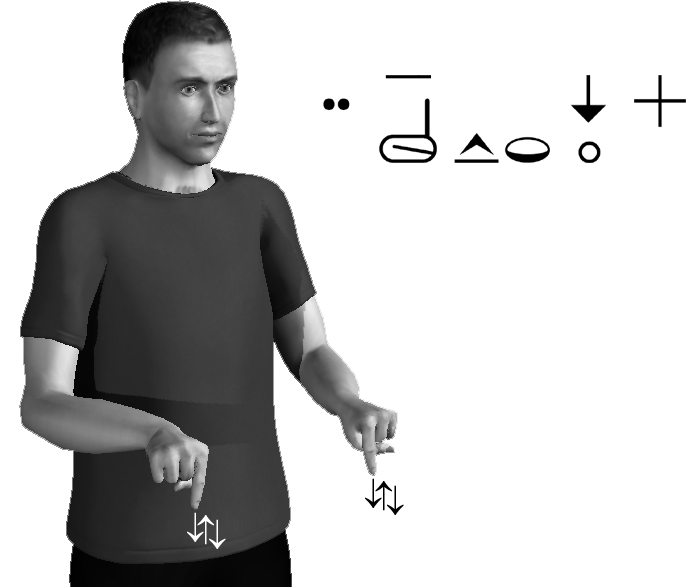}}
	\caption{British Sign Language for ``here'' with HamNoSys transcription}
	\label{fig:here-hns-avatar}
\end{figure}

Like most such notations, HamNoSys records signs in terms of hand shape, hand location, and hand movement. It uses a set of special glyphs comprising a HamNoSys font. An example is shown in Figure~\ref{fig:here-hns-avatar}.
The symbols mean:
\begin{description}
\item[{\resizebox{!}{3ex}{\includegraphics{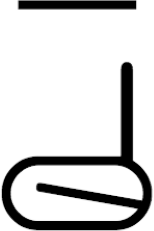}}}]
\item[\resizebox{!}{3ex}{\includegraphics{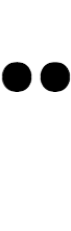}}]
The hands perform mirror-image actions.
\item[\resizebox{!}{3ex}{\includegraphics{images/here2hs.png}}]
The hand shape, illustrated in Figure~\ref{fig:modified-handshapes}, is the \textit{finger2} shape of Figure~\ref{fig:handshapes} with the thumb across the fingers and the index finger bent at the base.
\item[\resizebox{!}{3ex}{\includegraphics{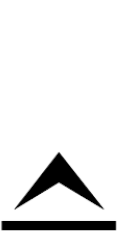}}]
The axis of the hand points forwards.
\item[\resizebox{!}{3ex}{\includegraphics{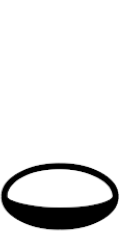}}]
The palm faces down.
\item[\resizebox{!}{3ex}{\includegraphics{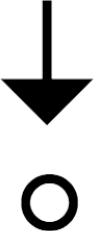}}]
The hands move down a small distance.
\item[\resizebox{!}{3ex}{\includegraphics{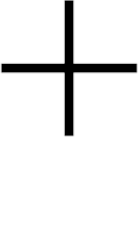}}]
Repeat the movement once from the initial posture.
\end{description}

In the most recent version, HamNoSys has been extended to cover facial movement and expression, and non-manual movements of the upper body such as nodding the head, raising the shoulders, etc.

\subsection{Principles of HamNoSys}

\begin{figure}
  \centering
  \resizebox{!}{5cm}{\includegraphics{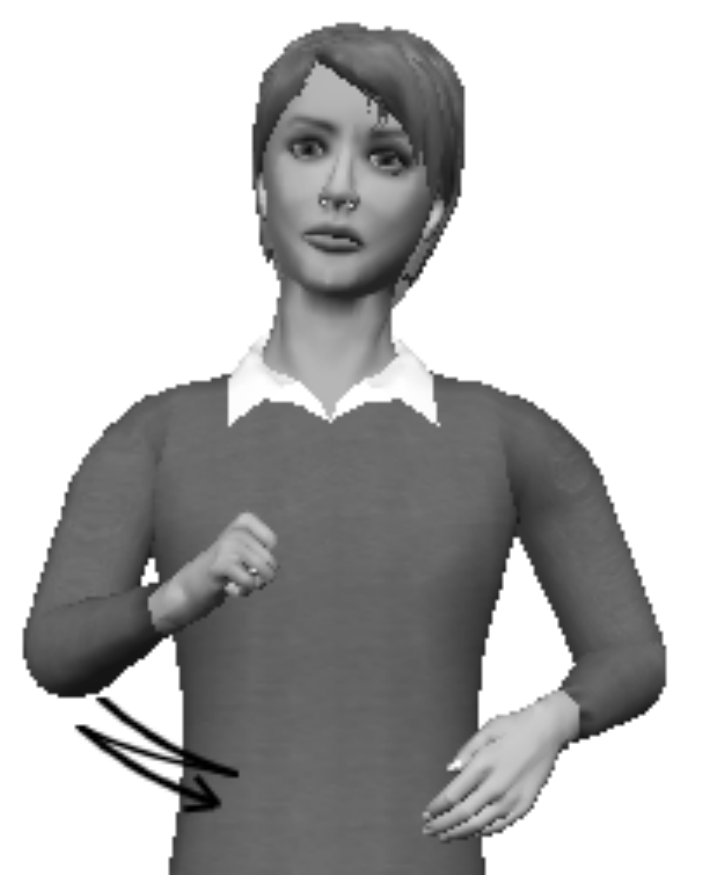}}
	\caption{BSL sign for ``Scotland''}
	\label{fig:ScotlandBSL}
\end{figure}

A HamNoSys transcription is independent of the person or avatar performing it. It records only those aspects of the gesture which are significant for the correct performance of a sign. Thus for most signs, it records only what the hands do; the rest of the body is assumed to do whatever is natural in order for the hands to perform the gesture. There are a few signs for which some other body part is significant, such as the BSL (British Sign Language) sign for ``Scotland'', in which the right elbow is the important part (see Figure~\ref{fig:ScotlandBSL}). 
In such cases, HamNoSys contains provisions for specifying the significant body part.

HamNoSys also includes some non-manual actions of the torso, shoulders, head, and eyes: shrugging the shoulders, tilting or turning the head, etc. There is also a repertoire of facial morphs of two types: visemes (visual equivalents of phonemes, the visible mouth movements used in speech) and other facial movements, such as closing the eyes or puffing the cheeks.
These are classified into a number of ``tiers'', each tier being the set of movements that affects one part of the body:
the shoulders, the torso, the head, the eyes, the facial expression, and the mouth.
For each tier a sequence of actions can be given, and actions for different tiers being performed simultaneously.
There are only limited capabilities for synchronising different tiers with each other or with the signing actions of the arms and hands.
Non-manuals are not as fully developed a part of HamNoSys as manual gesturing,
and play a correspondingly limited role in our implementation of HamNoSys.

\begin{figure}
  \centering
  \begin{tabular}{cc}
  \resizebox{6cm}{!}{\includegraphics{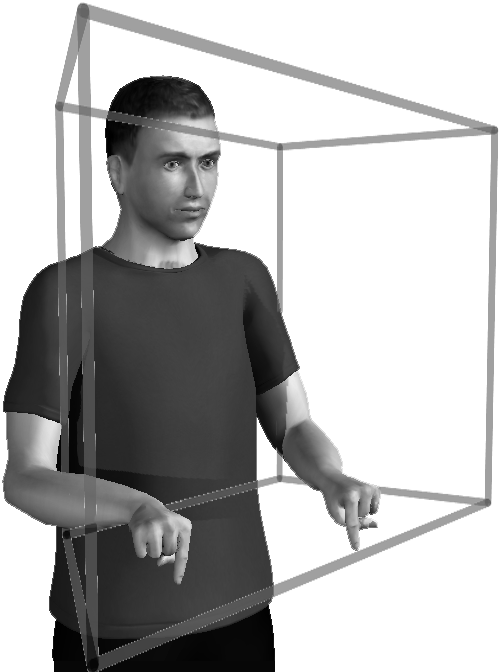}}
  &
  \resizebox{6cm}{!}{\includegraphics{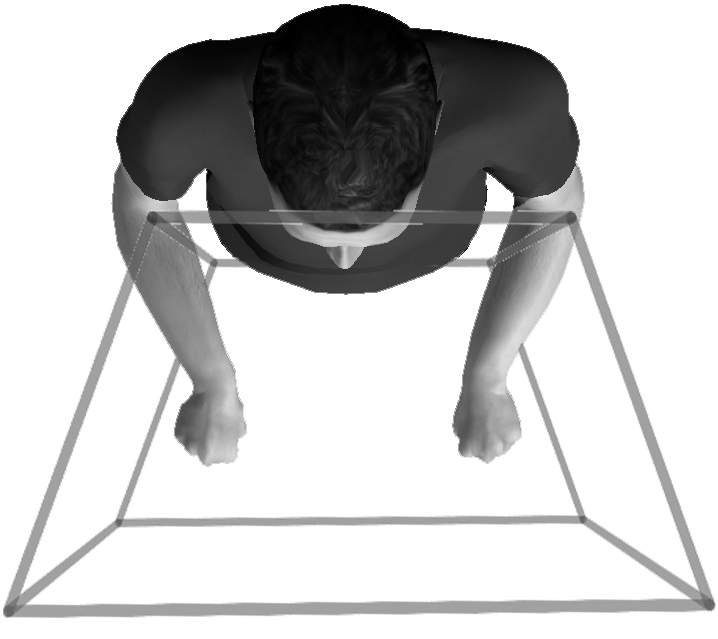}}
  \end{tabular}
	\caption{Signing space}
	\label{fig:signingspace}
\end{figure}

\begin{figure}
  \centering
  \resizebox{!}{5cm}{\includegraphics{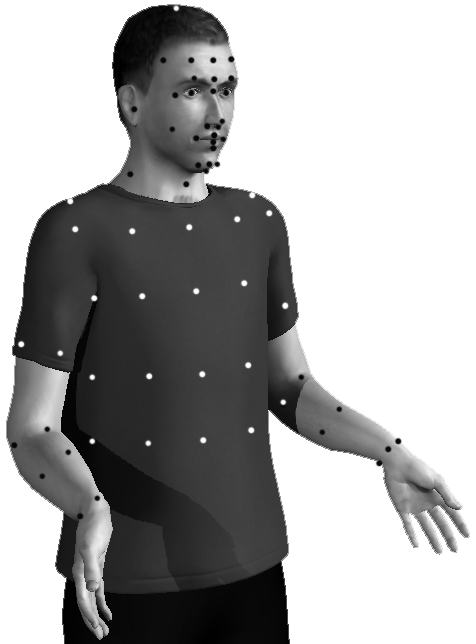}}
  \resizebox{!}{5cm}{\includegraphics{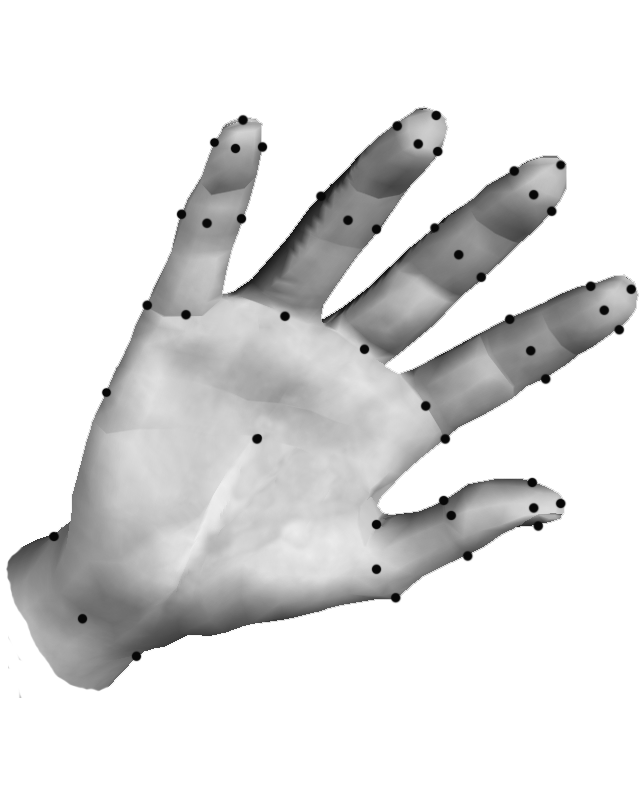}}
	\caption{Locations definable in HamNoSys on the surface of the body}
	\label{fig:locations}
\end{figure}

Positions are specified in terms of named locations.
These fall into two classes: locations in ``signing space'', the space generally in front of the signer's upper body, and locations on or close to the surface of the body (see Figures~\ref{fig:signingspace} and~\ref{fig:locations}).
Signing space is represented in HamNoSys by a three-dimensional grid of discrete points.
These are at a number of levels: below the stomach, stomach, chest, shoulders, neck, head, and above the head. From left to right there are five positions: far left, near left, centre, near right, and far right. There are three proximities: near, medium, and far.
Points on the body include the palms, the fingertips, the eyebrows, the nose, the chin, and many more; each may be specified with a proximity of medium, near, or touching. In total, there are some hundreds of locations that can be specified.
In addition, a position can be specified as the midpoint between two of the above positions.

\begin{figure}
\centering
\begin{tabular}{cc}
\resizebox{!}{2cm}{\includegraphics{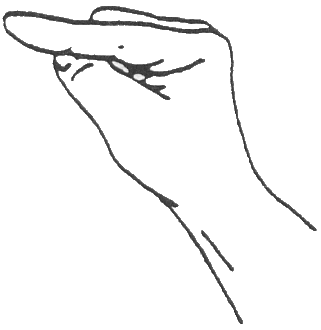}} &
\resizebox{!}{2cm}{\includegraphics{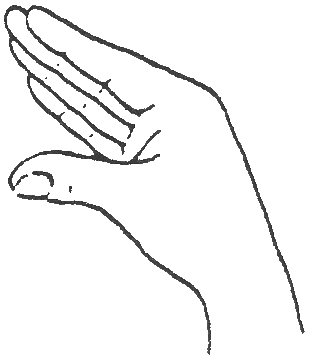}} \\
\resizebox{!}{0.7cm}{\includegraphics{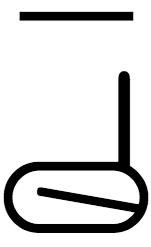}} &
\resizebox{!}{0.7cm}{\includegraphics{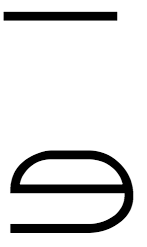}}
\end{tabular}
	\caption{Examples of modified hand shapes}
	\label{fig:modified-handshapes}
\end{figure}

Hand shapes are classified into the twelve basic types illustrated in Figure~\ref{fig:handshapes}.  Variations such as in Figure~\ref{fig:modified-handshapes} can be created by specifying the shapes of individual fingers: straight, rounded, curled, etc. Contacts between fingers can also be specified, such as fingers crossed, or thumb between fingers.
\newcommand{\showhandshape}[1]{
\begin{tabular}{c}
\resizebox{!}{2cm}{\includegraphics{images/hs-#1.png}} \\
\textit{\footnotesize #1}
\end{tabular}
}
\begin{figure}
\centering
\begin{tabular}{cccccc}
\showhandshape{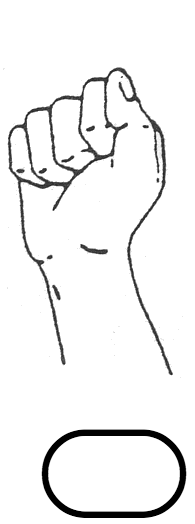} &
\showhandshape{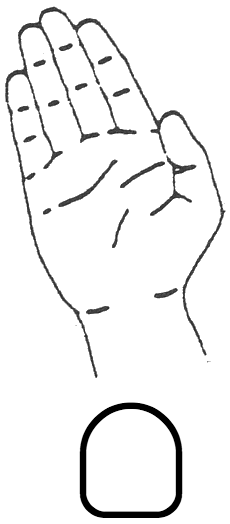} &
\showhandshape{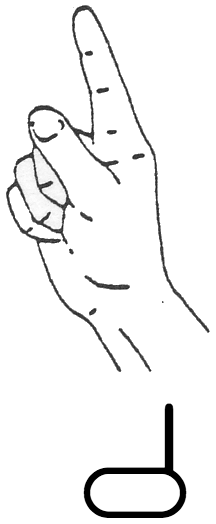} &
\showhandshape{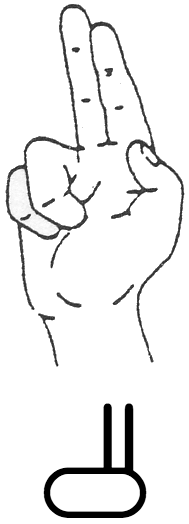} &
\showhandshape{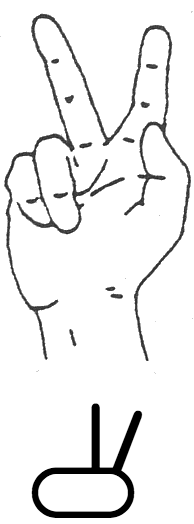} &
\showhandshape{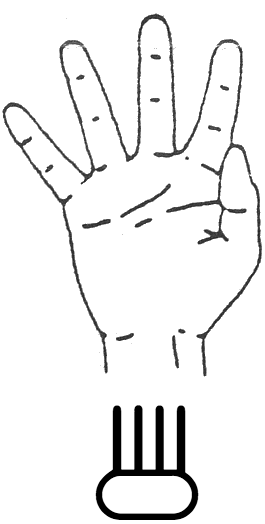}
\\
\showhandshape{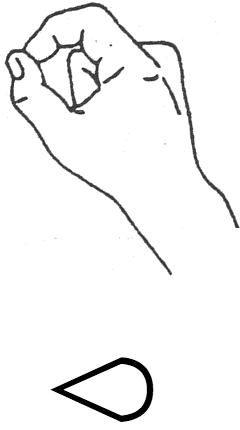} &
\showhandshape{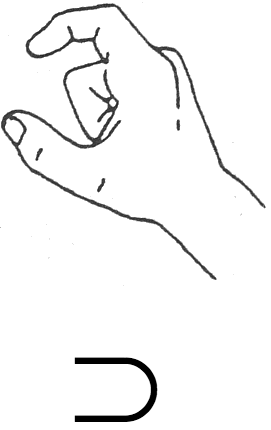} &
\showhandshape{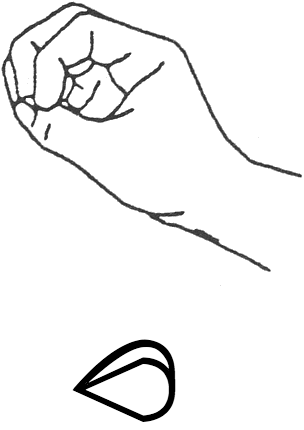} &
\showhandshape{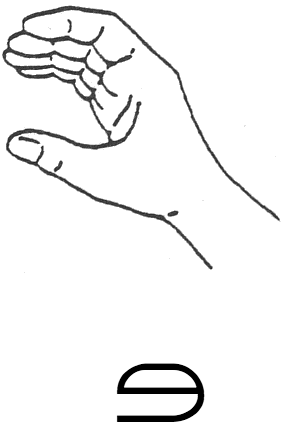} &
\showhandshape{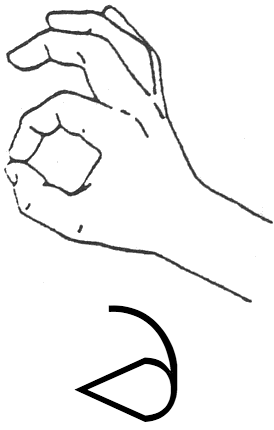} &
\showhandshape{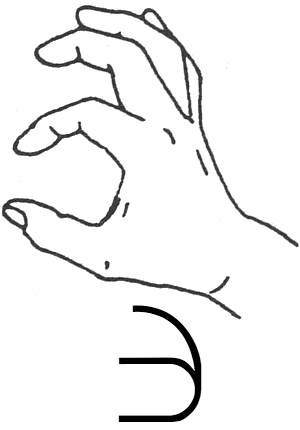}
\end{tabular}
	\caption{The basic hand shapes, their HamNoSys symbols, and their SiGML names}
	\label{fig:handshapes}
\end{figure}

Movements are parallel or sequential combinations of a repertoire of basic movements: straight, curved, and circular. The direction of a movement can be given as a combination of up, down, left, right, forwards, and back.
A target location can be given for a movement, instead of, or as well as, the direction of movement.

In many signs, the hands perform mirror image actions. HamNoSys allows transcriptions of such signs to be abbreviated, by specifying the action of the dominant hand, plus a symbol indicating that the nondominant hand mirrors the action. Left-right mirroring is the most common form, but there is also provision for front-back and up-down mirroring, as well as parallel movement of the two hands. The mirroring symbols can be regarded as abbreviations which could be eliminated by expanding the transcription.

For someone practiced in the use of HamNoSys, it takes about a minute to write down a transcription of a sign. This compares very favourably with the time it would take to either animate a sign by hand with animation software such as 3D Studio Max, or to record a sign in a motion capture session.

\subsection{SiGML}

HamNoSys was originally designed as a medium for researchers to communicate with each other, and not with computer processing in mind (other than storing HamNoSys transcriptions in databases).
As a result, it had initially no formal syntax, and only a verbal description of the semantics in the form of the HamNoSys manual~\cite{prillwitz.leven.ea:hamnosys}.
To separate out the issues of concrete syntax,
we obtained from the IDGS a first draft of a formal grammar, which was then further developed jointly.
The initial formal grammar was developed \textit{post hoc}, and in some cases was contradicted by actual practice in the existing corpus of HamNoSys transcriptions. Some transcriptions were indeed erroneous (as judged by the designers of HamNoSys), but others had to be accepted as correct and the grammar modified to allow them.

We then developed an XML application language to represent the grammatical structures of HamNoSys more abstractly, called SiGML: Signing Gesture Markup Language. A parser was written to convert HamNoSys into SiGML, and all subsequent processing is performed on the SiGML, which is read and parsed by the freeware \textit{expat} XML parser. Figure~\ref{fig:here-sigml} gives the SiGML translation of the HamNoSys of Figure~\ref{fig:here-hns-avatar}.
\newcommand{\codeindent}{\hspace*{1em}}
\begin{figure}
  \centering
\texttt{
\begin{tabular}{l}
<sign\_manual both\_hands="true" lr\_symm="true"> \\
\codeindent <handconfig handshape="finger2" thumbpos="across" mainbend="bent"/> \\
\codeindent <handconfig extfidir="o" palmor="d"/> \\
\codeindent <location\_bodyarm location="stomach" side="right\_at"/> \\
\codeindent <rpt\_motion repetition="fromstart"> \\
\codeindent \codeindent <directedmotion direction="d" size="small"/> \\
\codeindent </rpt\_motion> \\
</sign\_manual>
\end{tabular}}
\caption{SiGML transcription of ``here''}\label{fig:here-sigml}
\end{figure}

\section{Animating SiGML}

HamNoSys and SiGML are signer- and avatar-independent. To generate motion data for a particular avatar, numerical information about the avatar must be supplied, to be combined with the avatar-independent description of the movement, to produce avatar-specific motion data.
The information required divides into two classes, information about the geometry of a specific avatar, and information about body language or signing style, which can vary independently of the avatar.

The geometrical information required about the avatar consists of:
\begin{enumerate}
\item The dimensions of the avatar's skeleton:
the lengths of its bones, and the positions and orientations of the bones in some standard posture.
\item
For each of the non-manuals defined in HamNoSys,
the mixture of avatar-specific facial morphs or bone rotations which implements it.
\item
The coordinates of every location nameable in HamNoSys, relative to the bone which moves the relevant body part.
In principle, a HamNoSys location should be considered as a volume having a certain extent, the choice of which point within that volume to use being dependent on the context, but for implementation we have preferred to model each location by a single fixed point.
\end{enumerate}
The first of these can in principle be read automatically from whatever avatar description file is exported by the modelling software used to create it.
The second requires the avatar creator to express each of the HamNoSys facial movements as a combination of the avatar's morphs. 
The third requires the avatar creator to specify the coordinates of all the HamNoSys locations on the surface of the body. Some of these locations can be discovered automatically.  For example, it is fairly easy to determine surface points on the palmar, dorsal, ulnar, and radial sides of each finger, level with the midpoint of the middle phalanx. Other points are best placed manually: it is easier for the modeller to say where the ears are than to determine them algorithmically from analysis of the surface mesh.
Specifying the locations of these surface points should be considered a necessary part of the construction of any avatar intended for synthetic animation.
Note that no other knowledge is required of the surface mesh itself, which may be arbitrarily dense, subject to the requirement of rendering it at the desired frame rate.

The ``body language'' class of information consists of the mapping of HamNoSys categories to numbers. It requires:
\begin{enumerate}
\item
A numerical definition of the size of a ``large'', ``medium'', or ``small'' movement, ``near'' and ``far'' proximities, the shape of a ``deep'' or ``shallow'' curved arc, and similarly for all the other geometric categories in HamNoSys.
Distances are best given as proportions of various measurements of the avatar:
for example, ``far'' from the chest might be defined as a certain proportion of the length of either arm.
\item
A numerical specification of the time required for each type of movement, and of how that time should change when HamNoSys specifies the manner of movement as ``fast'', ``slow'', ``tense'', etc.
\item
A numerical specification of how the hands should accelerate and decelerate during movements of different types, performed with different manners.
\item
The temporal trajectories that the avatar-specific morphs and bone rotations that implement non-manual movements should follow. This takes the form of attack, sustain, and release times, plus a description of how the elements of the movement ramp up from zero to the full value during the attack, and down again to zero during the release.
\end{enumerate}
Body language information is specific, not to a particular avatar body, but to the personality inhabiting it.

Given all of this information, generating motion data requires solving the following problems.

\subsection{Hand location}
HamNoSys and SiGML always specify the locations of the hands, but to determine exactly what is meant by a piece of notation that is clear to the human reader can be surprisingly non-trivial.

The hand is an extended object: the whole hand cannot be placed at some point in space, only some particular point on the hand. HamNoSys allows a point on the hand to be specified, but does not require it. If it is omitted, an implementation of HamNoSys must guess which point is intended.  Generally, when the hands are not in contact with each other or the body, the centre of the palm is a reasonable choice, but this will usually not be suitable when there are contacts or close proxomities. For example, the BSL sign for ``me'' uses a pointing hand positioned close to or touching the chest, pointing to the signer.  If the centre of the palm is placed there, the index finger will penetrate the chest.  The significant location on the hand is here the index fingertip.

There is a trade-off here between the effort required of the program and its author in order to correctly guess the intended meaning, and the effort required of the writer of HamNoSys or SiGML in being explicit about such details.

In one significant case there is no way to express in HamNoSys certain information about hand location.  Some HamNoSys transcriptions express the spatial relationship of the two hands to each other, and the relationship of the pair of hands considered as a whole to the torso.  For the first part of this information, all that HamNoSys can express is the distance between the two hands, and the locations on the hands that are to be that distance apart.  The direction from one location to the other cannot be expressed in HamNoSys.  There are some heuristic rules that one can formulate for guessing this information, based on inspection of example transcriptions, but there is no way to be sure that the rules will be correct in all cases, because there is no formal definition of what the correct choice is.  For making new transcriptions directly in SiGML, it would be easy to extend SiGML with an additional attribute of the relevant XML element specifying the direction.  However, this would not help in dealing with legacy HamNoSys transcriptions, such as the IDGS corpus of several thousand signs of German Sign Language.

\subsection{Arm positioning}
Most HamNoSys transcriptions specify only the hand movements. The arms are assumed to move in whatever way is natural to achieve this.
The animation software must therefore have inverse kinematic rules to decide how the elbows and shoulders should move in order to place the hands into the required positions and orientations.
The rules that we have implemented were derived from informal observation of signers, experimentation, and by obtaining feedback from signers about the perceived quality of the postures and movement.

\begin{enumerate}
\item When a hand is in signing space, the plane formed by the upper and lower arm bones should make some angle $\alpha$ with the vertical plane through the shoulder and wrist joints.  This angle should vary from zero when the hand is directly to the right of the body, to some angle $\alpha_{\mbox{\textit{max}}}$ when the hand reaches across to the opposite side.
\item The wrist joint has limited mobility, especially for side-to-side movements.
Therefore the angle $\alpha$ chosen by the previous constraint may need to be further modified to prevent excessive wrist bending.
\item When the hand reaches beyond a certain distance from the shoulder, the collarbone should rotate to move the shoulder some distance in the direction of reach.
\item These rules are extended to cover all hand positions and orientations likely to arise in signing, in such as way as to make the resulting arm configuration vary continuously with the hand.
\end{enumerate}
For reaching movements across the body, the animation would be further improved by allowing the upper body to turn and tilt in the direction of reach, but this has not yet been implemented.

\subsection{Handshapes}\label{sect:handshapes}

HamNoSys handshape descriptions can be quite complex.
To a basic handshape selected from those of Figure~\ref{fig:handshapes} may be added any or all of the following:
\begin{description}
\item[\emdash]
A finger bending to override the bending of all the extended fingers or thumb of the basic handshape. The ``extended fingers'' are, for example, all of them for the \textit{flat} handshape, the index finger and thumb for the \textit{pinch12} handshape, etc. There are five possible bendings, beside the default value for the handshape, which can nominally be expressed in terms of the amount of bending of the three joints of the finger, as a proportion of the maximum possible. The joints are listed in order going outwards from the palm.
\textit{bent} $= (1,0,0)$,
\textit{round} $= (\frac{1}{2},\frac{1}{2},\frac{1}{2})$,
\textit{hooked} $= (0,1,1)$,
\textit{dblbent} $= (1,1,0)$,
\textit{dblhooked} $= (1,1,1)$.
The actual joint angles may vary substantially from these nominal values, and depend on the avatar.
\item[\emdash]
A thumb modifier: splayed, opposing the fingers, or folded across the palm.
In the case of the three \textit{cee}-type handshapes
(\textit{cee12}, \textit{ceeall}, and \textit{cee12open} in Figure~\ref{fig:handshapes}),
a thumb modifier may also specify that the gap between fingertips and thumb tip should be wider or narrower than normal.
\item[\emdash]
A specification of which fingers are to be the extended fingers: for example, to specify that the small finger should be extended instead of the index finger in the \textit{finger2} handshape.
\item[\emdash]
A specification that a particular finger should be given a specific bending.
\item[\emdash]
A specification that one finger should be crossed over another.
\item[\emdash]
A specification that the thumb should be placed between two fingers.
\end{description}
Turning a general handshape description into a set of joint angles appropriate to a given avatar would ideally be done by using the dimensions of the avatar's hands to calculate the joint rotations that would produce the correct relationships between all of its parts.
Instead of doing this, we calculated lookup tables for the avatars that we were using, by posing the avatar's hands in various shapes and reading the joint rotations. This involved less work, given the limited number of different avatars that we had available to animate, but in the longer run it would be preferable to calculate the tables automatically from the hand geometry.

The lookup tables express the basic handshapes and finger bending modifiers, not in terms of joint rotations directly, but in terms of a parameterisation of each joint according to the number of its degrees of freedom.  The middle and end joints of the fingers and thumbs are adequately described as hinge joints, and can be parameterised by a single number. The base joints are more complicated.

The base joint of each finger has two degrees of freedom, which can be labelled ``bend'' and ``splay''. In HamNoSys, these are independently specified. Bend is implied by the handshape or explicitly given by a fingerbending modifier. Splay is implied by the handshape: \textit{finger23spread}, \textit{finger2345}, \textit{cee12open}, and \textit{pinch12open} have various degrees of splay of the extended fingers, the other handshapes do not.
The problem is then to map any given amounts of bend and splay to the corresponding rotation, taking into account the interaction between the two.
The interaction has two effects.
Firstly, bend limits splay: when the base joint is fully bent towards the palm, it has almost no mobility about any other axis. So when the value of bend is at its maximum, the rotation should be independent of the value of splay.
Secondly, the space of all physiologically possible rotations of the joint is a two-dimensional subset of the three-dimensional space of all mathematically possible rotations. We require a description of that space which makes it easy to calculate the rotation corresponding to a given bend and splay. The third mathematically existing but physiologically infeasible degree of freedom is twist about the axis of the finger.

There are two standard types of mechanical joint having two degrees of rotational freedom: the universal joint and the constant velocity joint\footnote{The set of rotations of a constant velocity joint is the set of all rotations about all axes passing through a given point and lying in a given plane.
The name derives from its use in the front half-axles of a front wheel drive vehicle, where it ensures that when the driving side turns at a constant velocity, so does the driven side.}.
We found that neither of these, with their natural parameterisations, provides an accurate mapping of bend and splay to joint rotation, no matter how the bend and splay axes are selected. All of them require a significant amount of twist to be added for some combinations of bend and splay in order to reproduce the observed orientation of a real finger. To compute this twist one must either construct a lookup table defining twist as a function of bend and splay, deriving a continuous function by interpolation, or find an algorithmic method of computing the required total rotation of the joint.
We adopted the latter approach, and require just three extra measurements of each of the avatar's fingers. These could be automatically computed from the locations of some of the definable points on the hands, but at present have been measured manually for each avatar. The measurements are of deviations between the actual directions of the proximal and medial phalanges of the fingers in certain handshapes, and the directions they would have in an ideal simplified hand whose fingers in a flat handshape are straight and parallel to the axis of the hand, and which bend along axes exactly transverse to the fingers.
See Figure~\ref{fig:fingerangles}.
\newcommand{\flatsplay}{\theta_{\mbox{\scriptsize \textit{flat}}}}
\newcommand{\fistsplay}{\theta_{\mbox{\scriptsize \textit{fist1}}}}
\newcommand{\fisttwist}{\theta_{\mbox{\scriptsize \textit{fist2}}}}
\begin{figure}
  \centering
  \begin{tabular}{ccc}
  \resizebox{!}{3.5cm}{\includegraphics{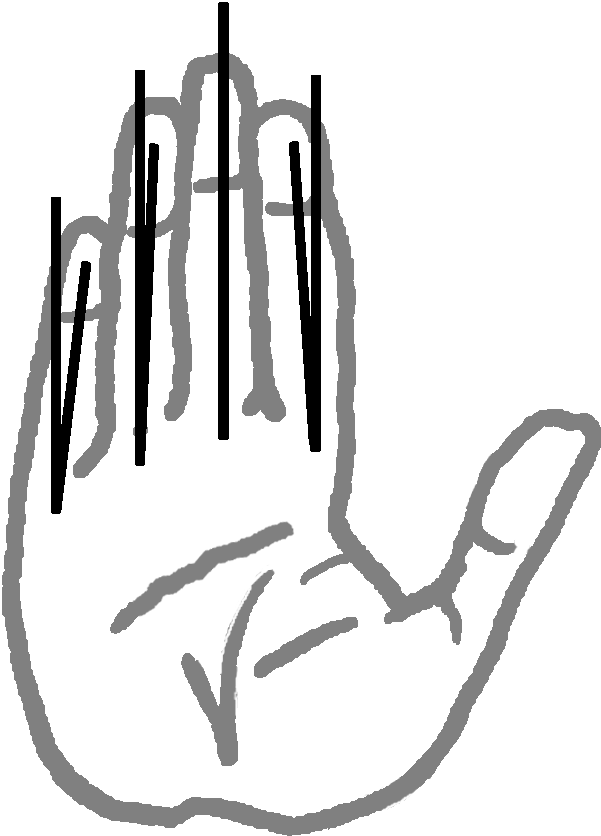}} &
	\resizebox{!}{2cm}{\includegraphics{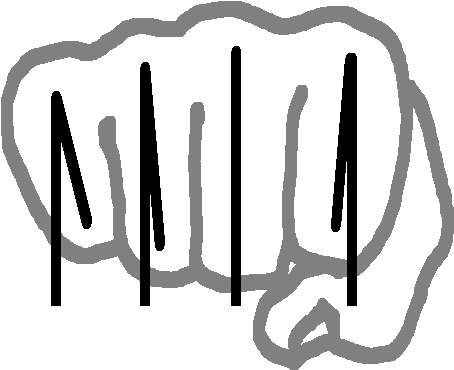}} &
	\resizebox{!}{2.5cm}{\includegraphics{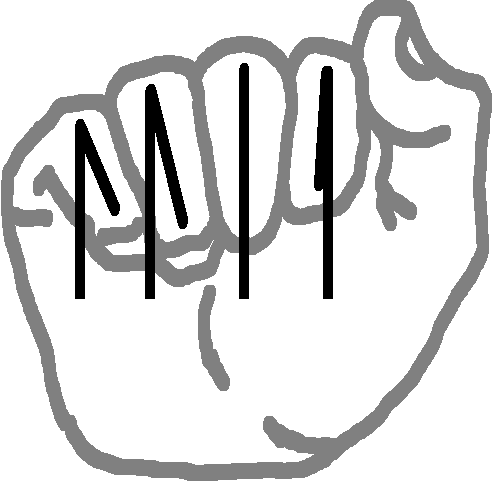}} \\
	$\flatsplay$ & $\fistsplay$ & $\fisttwist$
  \end{tabular}
	\caption{Angles characterising finger base joint articulation}
	\label{fig:fingerangles}
\end{figure}
\begin{enumerate}
\item
In the flat hand (in which the fingers are straight and touch each other all along their length), the angle $\flatsplay$ (a different angle for each finger) between the actual direction of the proximal phalanx of each finger and the ideal direction.
Typically, there is some inward tapering of the fingers, depending on their thickness.
\item
The same measurement, taken for a fist, giving angles $\fistsplay$ for each finger.
\item
The same measurement, taken for the medial phalanges in a fist, giving angles $\fisttwist$ for each finger.
\end{enumerate}
Let $\beta$, $\sigma$, and $\lambda$ be the bend, splay, and longitudinal axes of rotation. These are orthogonal: $\lambda$ is the axis of the hand, $\sigma$ is normal to the plane of the hand, and $\beta$ is perpendicular to both.
Write $(\alpha,a)$ for a rotation about axis $\alpha$ of angle $a$,
and define $\mbox{mid}(a,b,k) = a(1-k) + bk$.
Then given angles $b$ and $s$ for bend and splay, and defining $b' = 2b/\pi$, we construct the total rotation as the following composition of four components:
\[
(\sigma,\mbox{mid}(s,0,b')) \cdot
(\beta,b) \cdot
(\sigma,\mbox{mid}(\flatsplay,\fistsplay,b')) \cdot
(\lambda,\fisttwist b')
\]
When $b = 0$, this reduces to
$(\sigma,s+\flatsplay)$
and when $b = \frac{1}{2}\pi$, it reduces to
$(\beta,\frac{1}{2}\pi) \cdot (\sigma,\fistsplay) \cdot (\lambda,\fisttwist)$,
which can be verified to produce the finger orientations exhibited in Figure~\ref{fig:fingerangles} for the flat and fist handshapes.
(Some care is necessary about the choice of directions of the axes and signs of the rotation angles.)
In the figure, the three angles are all zero for the middle finger.
For $\flatsplay$, this is because we define the longitudinal axis of the hand to be the direction of the middle finger in the flat hand.
For $\fistsplay$ and $\fisttwist$, this is a contingent fact about the particular hand from which the illustrations were made.

\begin{figure}
  \centering
  \begin{tabular}{cccc}
\raisebox{1.15cm}{without:}
& \raisebox{0.25cm}{\resizebox{!}{2cm}{\includegraphics{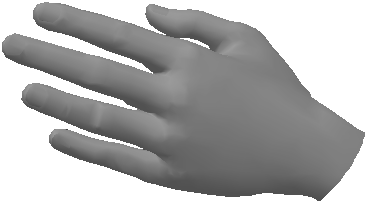}}}
& \raisebox{0cm}{\resizebox{!}{2.5cm}{\includegraphics{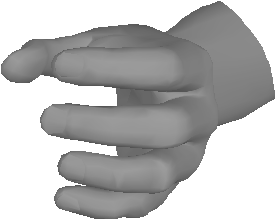}}}
& \raisebox{0cm}{\resizebox{!}{2.5cm}{\includegraphics{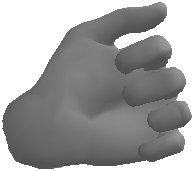}}}
\\
\raisebox{1.15cm}{with:}
& \raisebox{0.25cm}{\resizebox{!}{2cm}{\includegraphics{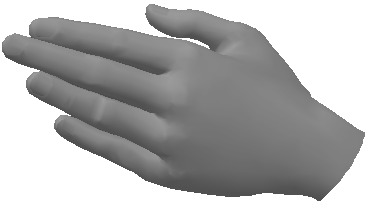}}}
& \raisebox{0cm}{\resizebox{!}{2.5cm}{\includegraphics{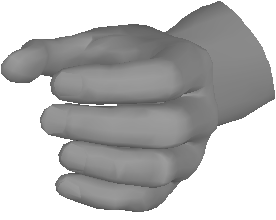}}}
& \raisebox{0cm}{\resizebox{!}{2.5cm}{\includegraphics{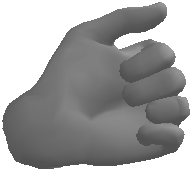}}}
\\
  & \textit{\footnotesize flat} & \textit{\footnotesize ceeall} & \textit{\footnotesize fist}
  \end{tabular}
	\caption{Handshapes implemented with and without the base joint articulation angles}
	\label{fig:claws}
\end{figure}

Although these angles are typically no more than a few degrees, they make a large difference to the naturalness of the resulting handshapes. Setting them all to zero produces unrealistic claw-like hands. Comparative examples are shown in Figure~\ref{fig:claws}. (The abnormal shortness of the medial phalanx of the ring finger is a defect of the avatar.)

The carpometacarpal joints of the fourth and fifth fingers (joints which lie inside the palm) have some limited amount of mobility which is used when forming a handshape in which the fingers wrap round an object such as a tennis ball. Although one is typically not aware of the movement, it is significant: without that mobility, one would hardly be able to pick up a tennis ball with the whole hand. These joints can be modelled as hinges with their axes lying obliquely across the palm, and which should receive nonzero rotations when the fourth and fifth fingers have nonzero values for both bend and splay.
(Many supposedly fully articulated avatars do not include these joints.)

The base joint of the thumb is more complicated. It appears to have two degrees of freedom, which can be very roughly characterised as the angle between the metacarpal bone and the axis of the hand, and the rotation of the thumb about that axis, but a mapping of two real parameters to the space of rotations that produces realistic articulation of all the required handshapes is difficult to synthesize. We eventually discarded the attempt to describe it in terms of bend, splay, and an algorithmically determined twist, and defined the necessary amount of twist for the basic thumb positions in the lookup tables describing the basic handshapes. Intermediate thumb positions are then defined by interpolation of these three parameters.

HamNoSys can also define a handshape as being intermediate between two given handshapes.  (An example is shown in Figure~\ref{fig:me}; the backslash is the betweenness operator.)
In this case we linearly interpolate between the representations of the handshapes in terms of joint parameters, before converting to joint rotations.

\subsection{Collisions}

A certain amount of collision detection and avoidance is required in order to prevent the avatar passing its arms or hands through its body.
However, we make the basic assumption that the transcription we are given does not describe something physically impossible, such as a hand movement that begins in contact with the chest and then moves inwards.
We have only found it necessary to implement collision avoidance for two situations:
\begin{enumerate}
\item When reaching across the body, the upper arm must be prevented from entering the torso, by suitably raising the elbow.
\item Because the front of the torso cannot be assumed to be a vertical plane, a movement that begins in contact with the torso and moves up, down, or across must be modified so as to avoid penetrating the surface of the torso.
\end{enumerate}
To fulfil these constraints, it is not necessary to know the complete surface mesh of the avatar, merely a sufficiency of locations on its surface for interpolation to yield a good enough result. The set of HamNoSys locations is sufficiently dense to achieve this. For this reason, the actual surface mesh of the avatar (which might be arbitrarily complex) does not need to appear in the avatar description used by the synthesis software.

\subsection{Trajectories}
The route by which a hand moves from one location to another is in general explicitly specified by the transcription, and therefore does not need to be decided by the software.
The accelerations and decelerations along that path are not specified in any detailed way. HamNoSys distinguishes five modalities of movement: normal, fast, slow, tense\footnote{``Tense'' here means a slow and steady movement, as if performed with great effort. This is visually quite distinct from a movement which is merely slow, even when the face (which may communicate some of the associated affect) is concealed.}, and ``sudden stop at the end'' (such as in a sign for ``punch''). In addition, we distinguish four separate types of ``normal'' movement in signing:
``targetted'', ``lax'', ``hard contact'', and ``linear''.
These are not explicitly notated, but are implied by context.

Targetted movement is used for meaningful movement, as opposed to lax movement, which is used for non-meaningful movements. There is a marked visual difference between the two.
The deceleration at the end of a targetted movement is much greater and shorter than the acceleration at the beginning, giving a velocity profile that peaks after the midpoint of the movement.
Lax movement has a broader velocity peak lying more in the middle of the movement, with a sharper initial acceleration and a longer final deceleration than targetted movement.
An example of lax movement is in the return stroke of most repeated movements, such as the upwards stroke of the sign for ``here'' in Figure~\ref{fig:here-hns-avatar},
and the outwards stroke of the elbow in Figure~\ref{fig:ScotlandBSL}.
HamNoSys can also notate a form of repetition in which the return stroke is as significant as the forward stroke; for a sign using this repetition, both directions would be performed with the targetted trajectory.

The distinction between lax and targetted movement can also be seen in the transitional movement from the end of one sign to the beginning of the next.
When the second sign consists of a single static posture (such as the BSL sign for ``me'', in which the hand points to the signer), the transition is targetted, but when the second sign contains movement, the movement into its initial posture from the previous sign is lax.
(In an early version of the implementation, we used targetted motion for all inter-sign transitions, and received comments from deaf people that those movements looked wrong: they appeared meaningful when they were not. Using lax transitions except before static signs eliminated this criticism.)

Hard contact is targetted movement which ends with the hand contacting a solid object such as the other hand or the torso, and which produces a more sudden stop than a targetted movement ending in mid-air. At the end of the movement, the hand remains in contact with the object for a perceptible time before moving again.

Linear movement, where the distance moved is proportional to elapsed time, is never appropriate for straight line movements in lifelike humanoid animation, but it is applicable when the hand makes a circular motion. Apart from the transitions into and out of the movement, the angular velocity remains close to constant.

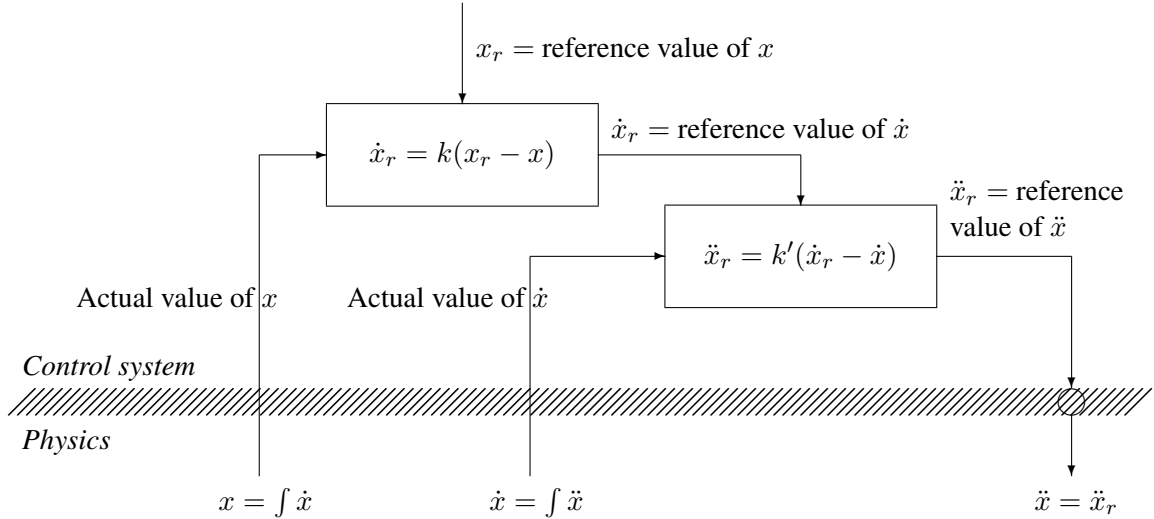
\begin{figure}
\newcommand{\figtext}{\normalsize}
	\centering
%	\resizebox{10cm}{!}{MISSING}%{\includegraphics{control.png}}
	\setlength{\unitlength}{0.9cm}
	\begin{picture}(15.5,8)(0,-0.7)
	  % path for x_r
	  \put(6,7){\vector(0,-1){1.5}}
	  \put(6.2,6.2){\figtext $x_r =$ reference value of $x$}
	  % path for x
	  \put(3,0){\line(0,1){4.75}}
	  \put(3,4.75){\vector(1,0){1}}
	  \put(0.3,2.5){\figtext Actual value of $x$}
	  \put(2.4,-0.5){\figtext $x = \int \dot{x}$}
	  % path for \dot{x}_r
	  \put(8,4.75){\line(1,0){3}}
	  \put(11,4.75){\vector(0,-1){0.75}}
	  \put(8.2,5){\figtext $\dot{x}_r =$ reference value of $\dot{x}$}
	  % path for \dot{x}
	  \put(7,0){\line(0,1){3.25}}
	  \put(7,3.25){\vector(1,0){2}}
	  \put(4.3,2.5){\figtext Actual value of $\dot{x}$}
	  \put(6.4,-0.5){\figtext $\dot{x} = \int \ddot{x}$}
	  % path for \ddot{x}_r
	  \put(13,3.25){\line(1,0){2}}
	  \put(15,3.25){\vector(0,-1){1.95}}
	  \put(15,1.1){\circle{0.4}}
	  \put(15,0.9){\vector(0,-1){0.9}}
	  \put(13.2,3.85){\parbox{3cm}{\figtext \flushleft $\ddot{x}_r =$ reference\\value of $\ddot{x}$}}
	  \put(14.45,-0.5){\figtext $\ddot{x} = \ddot{x}_r$}
	  \put(-0.5,1.5){\figtext \itshape Control system}
	  \put(-0.5,0.4){\figtext \itshape Physics}
		\multiput(-0.7,0.9)(0.133,0){125}{\line(1,1){0.391}}
		\put(4,4){\framebox(4,1.5){\figtext $\dot{x}_r = k(x_r - x)$}}
		\put(9,2.5){\framebox(4,1.5){\figtext $\ddot{x}_r = k'(\dot{x}_r - \dot{x})$}}
	\end{picture}
	\caption{Semi-abstract biocontrol model for synthesizing temporal trajectories}
	\label{fig:biocontrol}
\end{figure}

To synthesize suitable temporal trajectories for the different manners, we use the semi-abstract biocontrol model illustrated in Figure~\ref{fig:biocontrol}.
This is a two-level cascade control system built from two proportional controllers. The part below the shaded line is the simulated physics.
The controlled variable is $x$, representing the proportion of the distance that has been travelled towards the desired final position, and is initially zero.  $x_r$ is the reference or target value, initially 1.
The output of the outer controller is $\dot{x}_r = k(x_t - x)$, a demanded rate of change of $x$ which is proportional to the error, the difference between the reference value $x_r$ and the actual value $x$. This is the reference input to the inner controller, which compares it with the actual rate of change $\dot{x}$, and outputs a demanded acceleration $\ddot{x}_r$ proportional to the error $\dot{x}_t - \dot{x}$. The virtual actuator sets $\ddot{x}$ equal to $\ddot{x}_r$, which determines the time evolution of $\dot{x}$ and $x$.

\begin{figure}
	\centering \small
	\begin{tabular}{ccc}
	\includegraphics{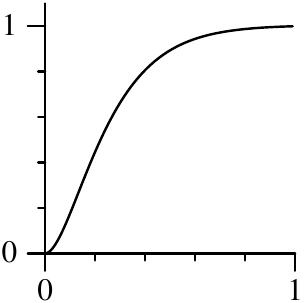} &
	\includegraphics{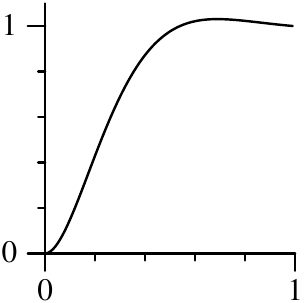} &
	\includegraphics{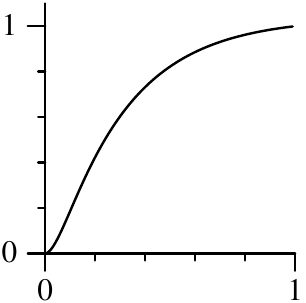} \\
	Critically damped ($k'/k=4$) & Underdamped ($k'/k=2$) & Overdamped ($k'/k=8$) \\
	\multicolumn{3}{c}{Trajectories of the basic biocontrol model.} \\
	\multicolumn{3}{c}{\ } \\
	& \resizebox{1.2in}{!}{\includegraphics{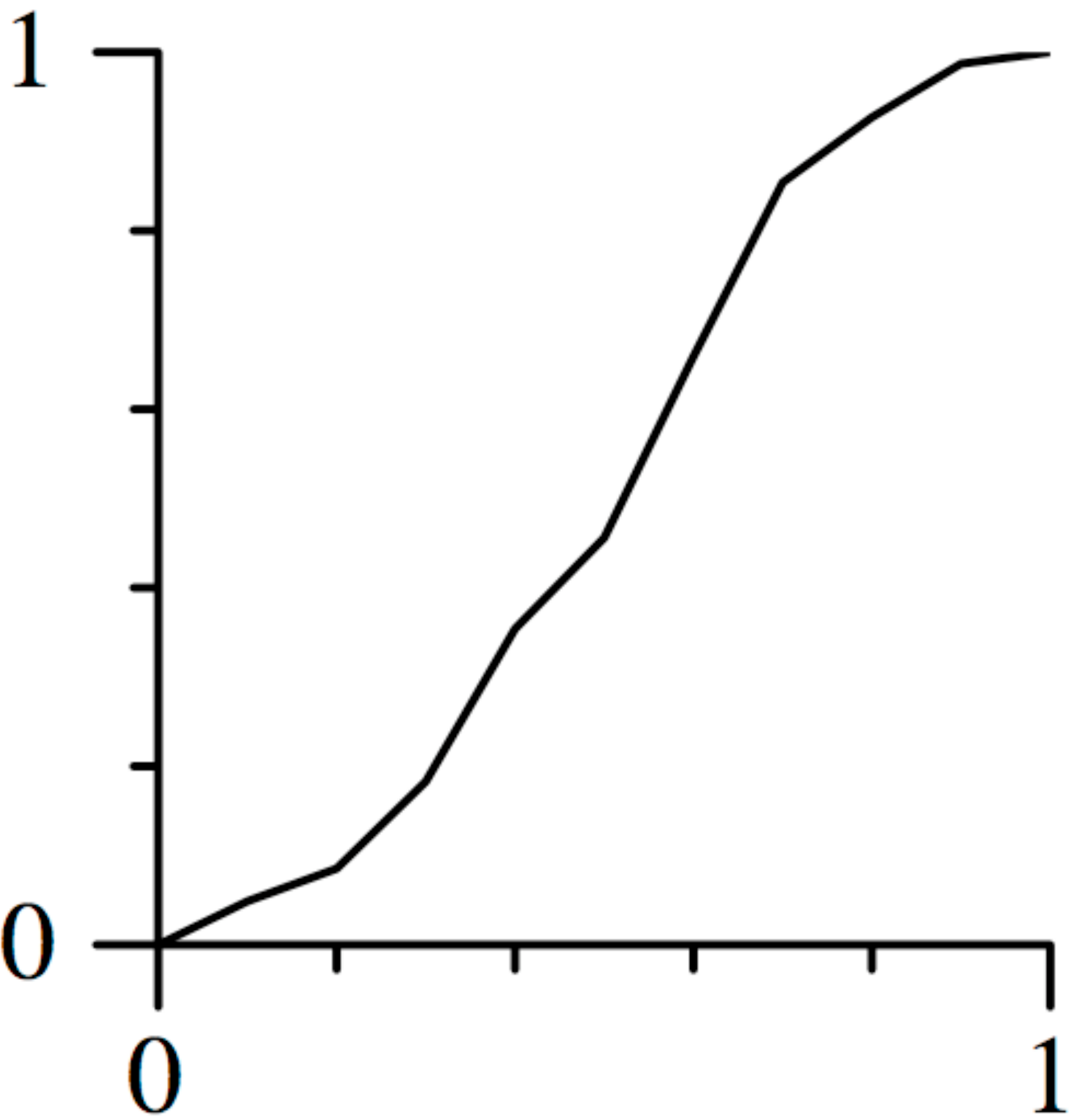}} & \\
	\multicolumn{3}{c}{Trajectory of human signer measured from video. Compare with ``targetted'' below.} \\
	\multicolumn{3}{c}{\ } \\
	\includegraphics{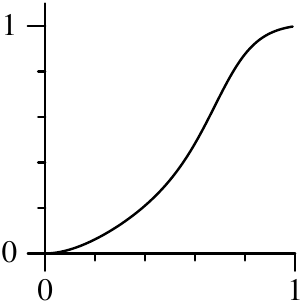} &
	\includegraphics{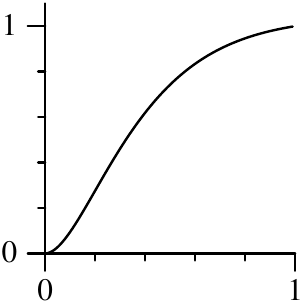} &
	\includegraphics{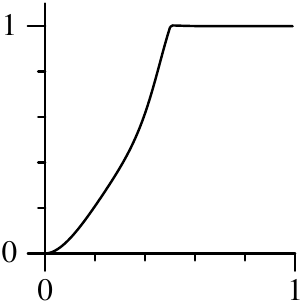} \\
	Targetted & Lax & Sudden stop \\
	\multicolumn{3}{c}{Trajectories of the modified biocontrol model.}
	\end{tabular}
	\caption{Horizontal axes: time; vertical axes: position.}
	\label{fig:trajectories}
\end{figure}

The equation of motion of the resulting system is
\[\ddot{x} + k'\dot{x} + kk'x = kk'x_t\]
With constant positive gains in both of the proportional controllers, this is mathematically equivalent to damped simple harmonic motion, critically damped when $k'/k = 4$, underdamped for smaller values of $k'/k$, and overdamped for larger values.
The top row of Figure~\ref{fig:trajectories} illustrates some trajectories of the system. The horizontal axis is scaled so that the major part of the motion (when it has settled to within 99\% of the final value) occurs in one unit of time.

We found that all choices of $k'/k$ gave sluggish behaviour, compared with the performance of signers, especially towards the end of the movement.
We modified the system by making the stiffness increase as the target is approached (that is, by increasing $k$ and $k'$ as the error $x_r - x$ decreases, while keeping their ratio constant), so as to model the increase in tension of the muscles as the intended location is reached. This resulted in animations that were judged to be satisfactory, and compared well with trajectories observed by frame-by-frame stepping through a signing video, of which an example is given in the middle row of Figure~\ref{fig:trajectories},
for the sign ``west'' (the flat right hand, facing left, makes a leftwards movement). The movement lasts for 11 frames at 15 frames per second.
The ``sudden stop'' modality uses a greater increase of $k$ and $k'$, sets their ratio to be underdamped, and truncates the trajectory when the final position is reached.
Varying the parameters of the system produced suitable trajectories for each of the other HamNoSys modalities, as well as the distinctions between targetted, hard contact, and lax movement.
Because HamNoSys manners are a discrete set, we can greatly reduce the computation time by precomputing a trajectory for each one, in the form of a lookup table for a function $f:[0\dots1]\rightarrow[0\dots1]$.
When a fraction $k$ of the duration of the movement has elapsed, the hand has moved a fraction $f(k)$ along its path.
The bottom row of Figure~\ref{fig:trajectories} illustrates some of these trajectories. The ``targetted'' trajectory is the one that would be used to animate the sign that was measured from video.

In our implementation, the time taken for a movement is independent of the size of the movement.
Some researchers have adjusted the timing according to Fitts' law~\cite{Fitts}, which states that the time for a reaching movement is a linear function of the log of the distance. However, his observations were for reaching movements to a target of definite size and location, in which subjects were asked to favour accuracy over speed.  Signing movements are not of this nature. For example, the movements of the hands in the BSL sign for ``here'' illustrated in Figure~\ref{fig:here-hns-avatar} have no particular target location of any definite size, and the sign for ``Scotland'' (Figure~\ref{fig:ScotlandBSL}) does not involve hand movements at all.
Observation of signers' movements suggests there is in fact little variation of duration with distance (which, at the level of accuracy and range of distances we are dealing with, is also a reasonable interpretation of the logarithmic relationship expressed by Fitts' law).

The duration can be observed to vary with the shape of the movement: curved arcs are performed more slowly than straight movements, and circular movements take longer still.
The durations used by the implementation were arrived at through feedback from signers.

We use the same temporal trajectories for all the other types of movement:
change of hand orientation, change of handshape, non-manual movements, and change of facial morphs.

\subsection{Interpolation between postures}
The trajectories defined in the previous section specify what proportion of the path from one posture to the next the avatar should have moved after any given elapsed time. We now describe the actual interpolation between postures.

The location of a hand at any instant is defined by the spatial path described by the SiGML, the particular point on the hand that must traverse this path, and the proportion of the path that has been traversed.  Interpolation is linear with respect to distance travelled along the path.

The path in rotation space from one hand orientation to the next is that defined by spherical linear interpolation.

The location and orientation of the hand at each moment define the location of the wrist joint.  From this, the wrist, elbow, shoulder, and clavicle joint rotations are calculated by inverse kinematic routines.

When the hand changes shape, the rotation of each hinge joint of the fingers and thumb is interpolated in the obvious way. The finger base joints interpolate the bend and splay parameters, and the thumb base joints the bend, splay, and twist parameters described in section~\ref{sect:handshapes}.

Non-manual movements and facial morphs are described as real numbers expressing the amount of each one that is to be present at any instant, and therefore linear interpolation applies.

\subsection{Performance}
Each frame of animation data consists of about 50 bone rotations, the exact number depending on the avatar: 15 to 17 joints in each hand, 4 in each arm, and several joints in the spine, neck and head. The frame also includes an amount of each of the avatar's facial morphs.

The current system is capable of generating 15,000 frames of animation data per second on a 1.7~GHz desktop PC.  This excludes the time taken to read and parse the SiGML input.  Although the current system does store SiGML as text, when embedded in a system such as those described in Section~\ref{section:otherwork}
it would be feasible to bypass parsing by directly constructing SiGML data structures.

If the animation is to play at 25 frames per second, which gives acceptable legibility for signing\footnote{A commercially produced CDrom for students of signing, from which the example trajectory in Figure~\ref{fig:trajectories} was measured, records all video clips at 15~fps.}, this means that the synthesis of motion data from SiGML data structures requires only 0.17\% of the time budget, even on what is by current standards a modestly powered machine.  At the 60~fps required for fast-action videogames, the fraction required of the time budget is still only 0.4\% per avatar.

Synthetic animation from a notation such as SiGML also allows for a great reduction in the amount of data to be stored. The SiGML displayed in Figure~\ref{fig:here-sigml} represents about 0.8 seconds of animation in about 320 bytes.  The corresponding 20 frames of 25~fps animation would (uncompressed) occupy about 12kB, assuming 50 rotations per frame, each represented as a triple of floating point numbers.  The latter figure can be substantially reduced by various compression schemes, but it is unlikely that any such scheme could achieve the 37-fold reduction obtained by the SiGML representation.  The SiGML itself is susceptible of compression: the HamNoSys in Figure~\ref{fig:here-hns-avatar} only takes 10 bytes.

\section{Automatic generation of SiGML}

SiGML is designed to be human-readable and -writable, but for some applications it is necessary to generate SiGML automatically. One example is its original application, sign language.

Signing is often conventionally transcribed as a sequence of ``glosses'': names of the signs performed, such as \signname{BOOK} \signname{GIVE}, corresponding to the English sentence ``Give me the book''.  However, this is misleadingly simplified: to animate such a sequence it is not sufficient to merely replace each gloss by its SiGML transcription and generate animation data. This is because many signs take different forms depending on their context.  For example, \signname{GIVE} will be made with a handshape indicating the type of object that is being given, and the start and end locations of the movement will indicate who is giving to whom: you to me, him to her, me to multiple people, etc. \signname{GIVE} requires a set of parameters: the above example is better glossed as
\signname{BOOK} \signname{GIVE}(\signname{you,me,flat-thing}).  This variety of instantiations greatly expands the corpus of signs required for interactive discourse beyond what a mere count of the glosses in a dictionary might suggest. For highly inflected signs such as \signname{GIVE}(\dots), the SiGML transcriptions of its required forms must be generated automatically.
Even for signs which might not appear to require parameterisation, such as signs for ordinary concrete objects like \signname{BOOK}, the location at which the sign is performed may express some of the meaning.
For example, it may be performed at some arbitrarily chosen location in signing space, in order to be able to refer back to the object later in the discourse by pointing to that location.

In the ViSiCAST project, sign language has been generated from English text via the intermediate form of Discourse Representation Structures and Head-Driven Phrase Structure Grammars.
The latter form the input to sign language generation based on the Attribute Logic Engine. This pathway has been described in~\cite{Saf01b,Saf02,MarS04}, and will not be further described here.
Sign language represented in the parameterised form described above can then be transformed into SiGML, inserting appropriate parameters such as handshape and location.

In virtual reality applications with autonomous interactive avatars, SiGML (or an extension of SiGML covering a wider repertoire of movements) could be generated from the system's internal high-level representation of the avatars' intended actions (e.g.~``point to that building''). The SiGML description serves as a low-level, but still avatar-independent, representation of the desired action, which can then be translated into motion data for the particular avatar.

\section{Limitations and extensions of HamNoSys}

The success of our implementation of SiGML demonstrates the practicality of real-time synthesis of human movement in a certain restricted setting.
We are interested in expanding this application to wider repertoires of human movement, such as are required by general interactive virtual reality applications.
The experience of implementing SiGML has shown some weaknesses of HamNoSys for this application, and suggests ways in which it might be improved.

\subsection{Limitations}

\subsubsection{Syntax}

HamNoSys was originally developed for people to read and write. Its formal grammar was developed in hindsight to fit existing practice. While that practice is easy for those trained in the notation to deal with, the resulting grammar is more complicated than is warranted by the subject matter. This is still true of SiGML. Although it eliminates many of the irrelevant complexities of the concrete syntax of HamNoSys, it is still very much based on HamNoSys. A notation developed specifically for computer animation could be substantially simplified without any loss of expressiveness.

\subsubsection{Semantics}

HamNoSys depends to a significant extent on human understanding of signing
for both parsing and interpretation. A large part of the task of implementing SiGML was taken up by attempts to automatically fill in information that human transcribers typically omit when transcribing signs, and (often unconsciously) fill in when reading transcriptions.

In the eSign project, among the people who were responsible for creating SiGML transcriptions of the signs required for various applications, those who had less prior experience with HamNoSys sometimes found it easier to produce transcriptions that worked well with the animation system, being less accustomed to depending on human interpretation.

An improved SiGML would have a simpler and more precise semantics, expressed in some form more explicit than merely its implementation.

\subsubsection{Handshapes}

The twelve basic handshapes of HamNoSys can, in principle, be mapped to (avatar-dependent) tables of finger and thumb joint angles, as can each of the fingerbending modifiers.
However, there turns out to be some significant context-dependency. For example, the actual bend angle of a finger that is ``bent at the base joint'' depends on the basic handshape
(compare the base joint angles for
the two handshapes
in Figure~\ref{fig:modified-handshapes}, which both use that finger shape modifier).
This requires a significant expansion of the per-avatar lookup tables used to translate handshapes to joint rotations. To handle this in a more flexible way, and to cover the less commonly used features of HamNoSys hand descriptions (such as crossed fingers), we augmented the SiGML \xe{handconfig} element with the possibility to explicitly specify finger joint angles.
This is, however, at the expense of avatar-independence; a better solution would be to calculate the required angles from an avatar-independent specification of the significant properties of the handshape.
Such properties would include contacts between different parts of the hand, or opposition of the thumb to specified fingers.

We also note that in the \textit{cee} handshapes, modifications expressing a wider or narrower gap between the fingers and thumb are expressed in HamNoSys by a modifying symbol which describes the thumb position. However, inspection of actual performance shows that changes to the finger-thumb distance are accomplished primarily by varying the bend angle at the base joints of the fingers. The HamNoSys notation for these handshapes does not map in a simple way to physical reality.

\subsubsection{Integration of manual and non-manual movements}

For historical reasons, HamNoSys makes an artificial distinction between manual and non-manual movements. This separation, enforced by the syntax of HamNoSys and carried over into SiGML, limits the flexibility with which the two classes of movement can be combined and synchronised with each other.

\subsection{Extensions}

\subsubsection{Discrete categories}

Several concepts of SiGML are discrete categories:
there is a finite set of locations, a finite set of distances, a finite set of directions and sizes of movement, etc. Other applications of scripted animation will require more precision and variability than these provide.
For some of these categories HamNoSys allows the specification of a value midway between two built-in values, but this still allows only a limited set of values.
Each of these sets of discrete values should be extended to continuous categories by adding the possibility of specifying fractional interpolation.
We have already implemented extensions to SiGML for this purpose.

\subsubsection{Timing}

We have provided more detailed control of timing than the ``fast'' and ``slow'' modalities of HamNoSys, by allowing an absolute duration or a temporal scaling to be applied to any movement or to a whole sign.

\subsubsection{Path specification}

A SiGML transcription specifies explicitly the spatial paths of
the hands --- straight line, curved, circular, etc. The
software never has to decide what path to follow to move a hand from one place to another. This should remain true for more general animation. The person or automated process that is composing a SiGML transcription may be assumed to be describing physically possible postures and movements. If the author has provided sufficient detail to uniquely determine the posture and movement, the software need not be burdened with the task of guessing or inventing missing detail.
If it is desired to ease the animator's task by allowing some details to be omitted and automatically synthesized, algorithms for doing so should be implemented either as a separate SiGML-to-SiGML translation or an external library that can be referenced from SiGML, rather than being built into the implementation of SiGML itself.

\subsubsection{Gravitational dynamics}

Signing movements are largely independent of gravity.
Hands move along their intended paths
and the arms do whatever is necessary to achieve that.  As a result, sufficiently lifelike accelerations and decelerations can be created from a simple biocontrol model with a small number of parameters, without requiring a detailed physical simulation.
In contrast, for lower body movements like walking, or for upper body movements that involve lifting heavy things, gravitational
forces are comparable with or greater than muscular forces and play a major
role in the dynamics, and hence also the visual appearance.
An extension of SiGML to cover these types of movement will require the implementation to be provided with a means of synthesizing them, which will be more complicated than the simple biocontrol model that we have found sufficient for signing animation.
Integration with more sophisticated methods such as those surveyed in~\cite{multon.france.ea:computer} would be desirable.

\subsubsection{Expressive movement}

HamNoSys was originally devised for recording signing citation forms,
and as such it contains only a discrete range of modalities, those necessary to the correct performance of the signs.
Thus SiGML does not have to deal with the entire range of
expressive movement.  Although signing itself can be very
expressive, the ``newsreader'' type applications in the eSign
project do not require it. However, it is an essential aspect of general character animation.

We regard expressiveness as being a concept at a higher level than the intended target of SiGML or any extension of it, and outside the intended scope of this work. SiGML is concerned with the avatar-independent description of concrete posture and movement; it is the responsibility of some higher-level system (such as, for example, some of the projects briefly surveyed in section~\ref{section:otherwork}) to express qualities of expressiveness in concrete terms.

\subsection{Principles of posture and movement notation}

Some of the principles underlying HamNoSys deserve wider application than they in fact receive in that notation.

\subsubsection{Constraint-based posture specification}

HamNoSys specifies the positions of the hands
either by their relation to each other,
and the relationship of the pair of hands considered as a single entity to the body,
or by the relationship of each hand to the body separately.
This is a useful method of defining posture in an avatar-independent way, but in HamNoSys it is subject to heavy restrictions.
The relationship is always one of distance only, not direction (which must be guessed from context). It applies only to position, with hand orientation always defined in absolute terms.
We are therefore considering a radically revised language which would specify postures in terms of constraints on the relative positions and orientations of general body parts.

\subsubsection{Intention vs.~geometry}

There is a large number of geometric features of a posture which a transcription notation might record, and there are many different ways of choosing a set of geometric features whose specification will completely determine the posture. For example, any two of the orientation of a hand, the orientation of a straight index finger, and the angle at the base joint of that finger determine the third. Which properties, therefore, should a notation record?

HamNoSys makes a particular choice of features to record.
For example, the direction of the hand is expressed in terms of the ``extended finger direction'' or e.f.d.: the direction the fingers would be pointing in if they were straight, or equivalently, the direction of the metacarpal bone of the middle finger.  However, there is evidence~\cite{crasborn:phonetic} that this property of the hand is not phonologically significant for signing. For a handshape that is pointing with the index finger, what is significant is the direction in which the index finger is pointing, which may be identified with the direction of the distal phalanx. So long as that direction is correct, the angle at the base joint of the index finger and the e.f.d.~of the hand can both vary considerably without affecting the meaning of the gesture.

\begin{figure}
  \centering
  \begin{tabular}{ccc}
	  \resizebox{!}{5cm}{\includegraphics{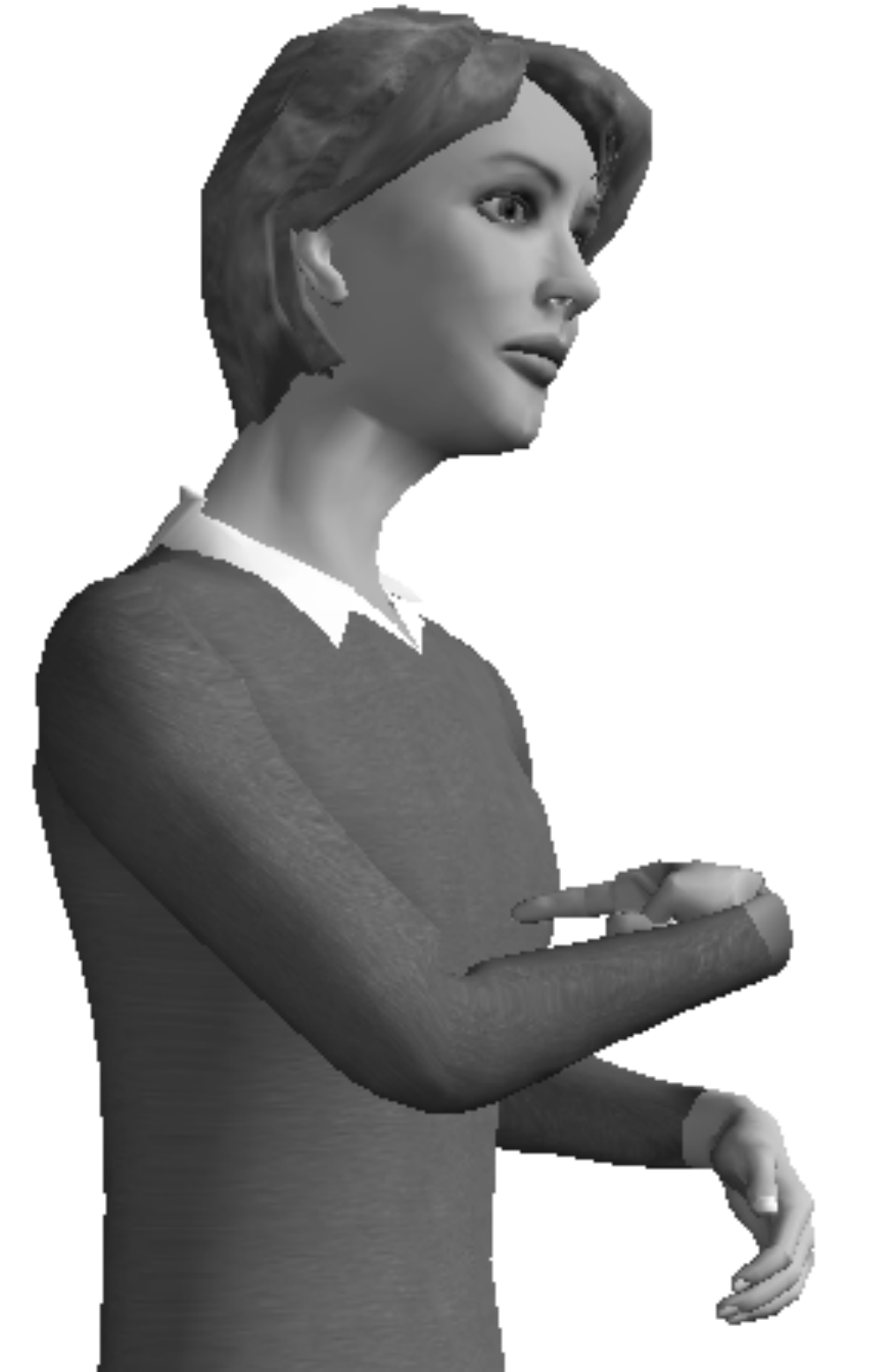}} & &
	  \resizebox{!}{5cm}{\includegraphics{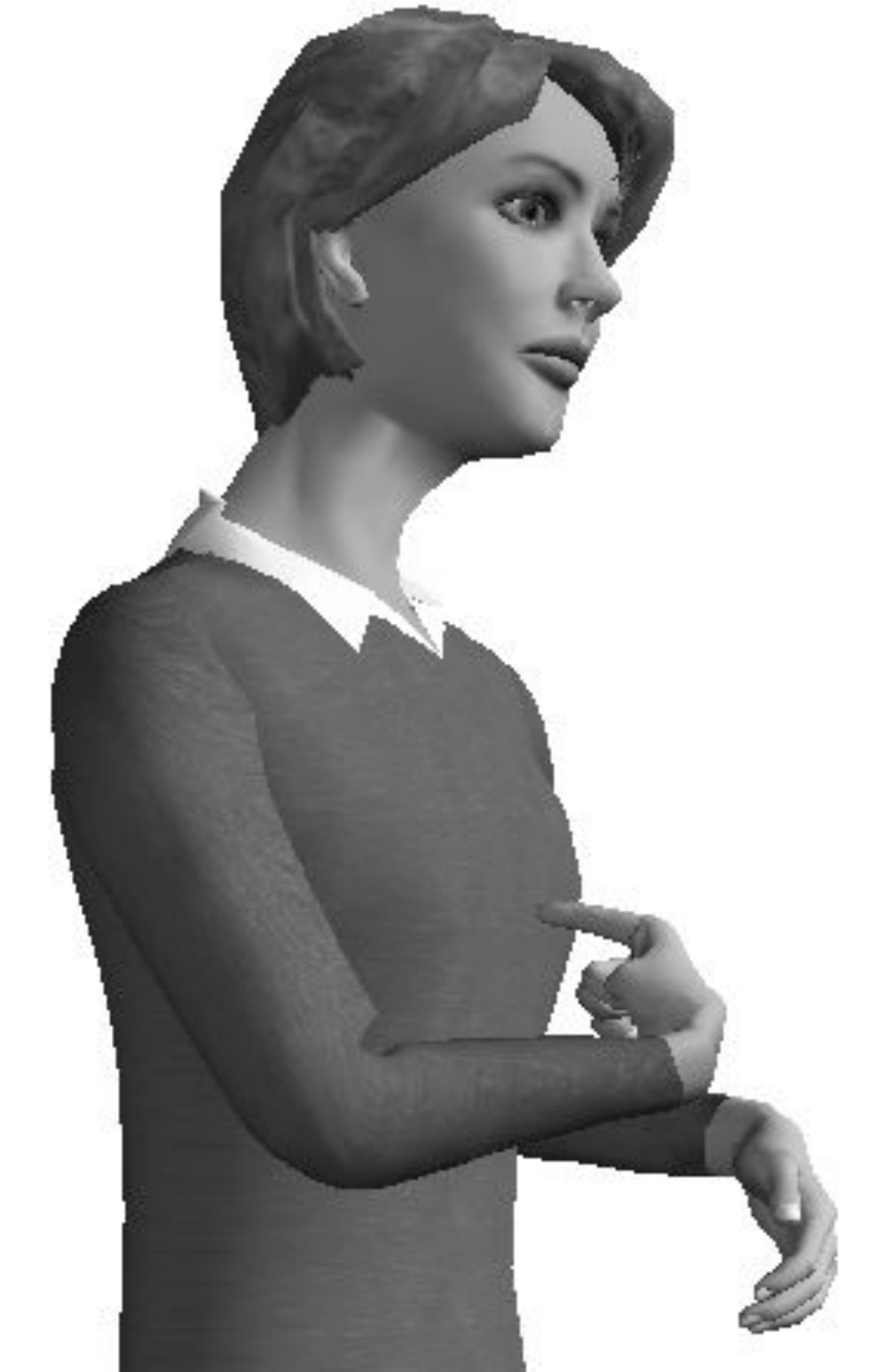}} \\
	  \resizebox{!}{0.7cm}{\includegraphics{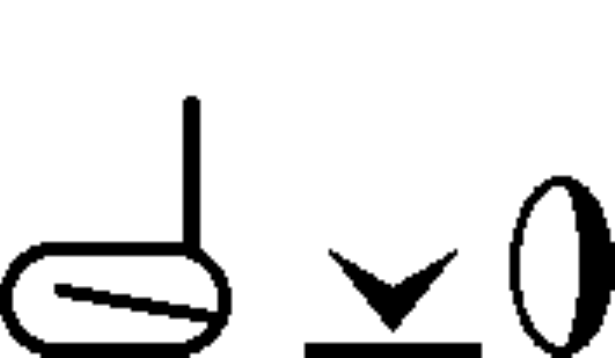}} & &
	  \resizebox{!}{0.7cm}{\includegraphics{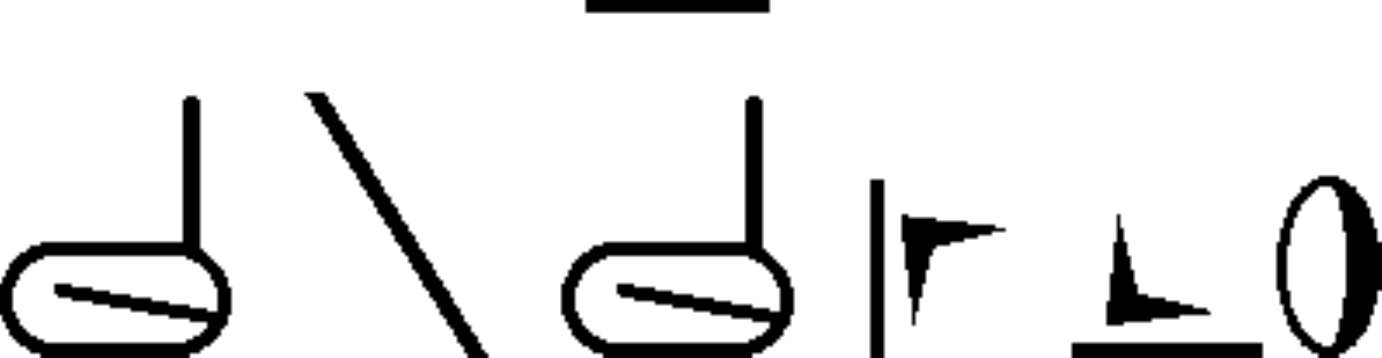}}
  \end{tabular}
	\caption{Incorrect and correct transcriptions of ``me''}
	\label{fig:me}
\end{figure}

This leads to a tendency in some circumstances for HamNoSys transcribers to record an idea of the sign which differs from its actual performance. An example is the sign for ``me'': an inward pointing index finger. This is commonly but incorrectly transcribed as a \textit{finger2} handshape with an inwards e.f.d. That would be a rather awkward posture to perform, and differs from actual performance. The transcription and actual performance are compared in Figure~\ref{fig:me}.

We therefore believe, as argued in~\cite{kennaway:experience}, that a gesture transcription notation should record those properties of a gesture which are essential to its significance: the properties which will remain invariant when the same gesture is performed by avatars of different proportions or different gestural styles.  In a slogan, \textit{intention is primary, geometry is secondary}. For a pointing handshape, the direction of pointing is significant; this is what should be recorded, and the bending of the pointing finger in Figure~\ref{fig:me} should be determined from inverse kinematic rules describing the performance of pointing gestures.

\section{Conclusions}

The present work has demonstrated the practicality of generating gesture animation in real time from an avatar-independent scripting language.
The generation of motion data is done not merely in real time, but in a very small fraction of the time budget, each single frame of data taking on average 1/15 of a millisecond to generate on a typical machine.  The animations are of a degree of naturalism which has proved acceptable to deaf users, in terms of both legibility and realism.
The SiGML representation of a movement is also many times smaller than the motion data which it generates.

Avatars for use with this method must be furnished with additional data beyond the skeleton topology and dimensions, and repertoire of morphs.  The additional data consists primarily of the locations of a standard list of named points on the surface of the body.

The work has suggested directions in which SiGML can be improved and extended in order to apply the method to other applications of real-time, naturalistic, humanoid animation.

\section{Acknowledgements}

We acknowledge funding from the European Union for our work on the ViSiCAST and eSign projects under the Framework V programme,
the support of our colleagues and partners on these projects, and the deaf users of sign language who have provided valuable feedback on the quality of our animations.
John Glauert and Ralph Elliott at the University of East Anglia collaborated with the author in designing the SiGML DTD,
and Ralph Elliott wrote the HamNoSys parser and the translator to SiGML.
Televirtual Ltd.,
one of our partners in ViSiCAST and eSign, produced the VGuido and Visia avatars, whose images appear in this paper; Figure~\ref{fig:claws} uses an avatar created by Vincent Jennings at the University of East Anglia.

\bibliography{animgen}

%{\itshape
%References still uncited:
%\begin{itemize}
%\item Pelachaud~\cite{CarCP02}
%\item Hodgins~\cite{hodgins.wooten.ea:animating}
%\item Unuma et al.~\cite{unuma.anjyo.ea:fourier}
%\end{itemize}
%}

\end{document}